\documentstyle[epsfig,nolta]{article} 

\newcommand{\Ss}{{\bf s}}
\newcommand{\xx}{{\bf x}}
\newcommand{\yy}{{\bf y}}
\newcommand{\Eepsilon}{{\mbox{\boldmath $\epsilon$}}}
\newcommand{\Eeta}{{\mbox{\boldmath $\eta$}}}
\newcommand{\av}[1]{\langle #1 \rangle}

\begin{document} 
\title{NONLINEAR PROJECTIVE FILTERING I:\\
BACKGROUND IN CHAOS THEORY}
\author{Holger Kantz$^{\dag}$ and Thomas Schreiber$^{\ddag}$}
\affiliation{\dag Max Planck Institute for Physics of Complex Systems,
      N\"othnitzer Str. 38, D--01187 Dresden\\
   {\tt kantz@mpipks-dresden.mpg.de}\\
   \ddag Physics Department, University of Wuppertal, 
   D--42097 Wuppertal\\
   {\tt schreibe@theorie.physik.uni-wuppertal.de}}

\maketitle
\abstract
We derive a locally projective noise reduction scheme for nonlinear time series
using concepts from deterministic dynamical systems, or chaos theory. We will
demonstrate its effectiveness with an example with known deterministic dynamics
and discuss methods for the verification of the results in the case of an
unknown deterministic system.
\endabstract

\section{INTRODUCTION}

The paradigm of deterministic chaos as an alternative explanation for complex
temporal behaviour~\cite{KantzSchreiber} has made the development of novel
signal processing techniques necessary. Deterministic chaotic systems are
characterised by exponentially decaying correlations and thus broad band power
spectra. Thus, except for the case of highly oversampled, time continuous
signals, filtering by frequency cannot be applied since signal and noise have
similar spectral properties. On the other hand, deterministic dynamics of the
form
\begin{equation}\label{eq:determ}
   \xx_{n+1}=F(\xx_n)
\,,\end{equation}
or given by an ordinary differential equation of first order satisfying a
Lipschitz condition for unique solutions, generates strong signatures when
viewed in its phase space. Nonlinear noise reduction methods have
been developed to exploit these structures. Conceptual as well as technical
issues arising in such a situation have been well discussed in the literature,
see Kostelich and Schreiber~\cite{noiserev} for a review containing the
relevant references.

In engineering applications, we usually face a different situation --- the
signals themselves often contain a stochastic component and we cannot make the
assumption that the system is of the form~(\ref{eq:determ}) and deterministic
chaos is present. It turns out that at least for a subclass of the nonlinear
filtering schemes, the phase space projection techniques~\cite{on},
deterministic chaos is not a necessary requirement. The only assumption that
has to be made is that the signal of interest is approximately described by a
manifold that has a lower dimension than some phase space it is embedded
in. This is formally true for low dimensional deterministic signals, but also
for certain stochastically driven nonlinear phenomena.

In this paper, a phase space projection scheme for noise reduction will be
motivated and described. We will give an example with known deterministic
dynamics for the purpose of illustration. Applications to real data are
discussed in Ref.~\cite{II} in this volume.

\section{METHOD}
Let $\{\xx_n\}$ be the states of a system at times $n=1,\ldots,N$, represented
in some vector space ${\cal R}^m$.  A $(m-Q)$--dimensional submanifold ${\cal
F}$ of this space can be specified by $F_q(\yy)=0,\quad q=1,\ldots,Q$. 
Suppose such a manifold exists such that the sequence of vectors $\{\xx_n\}$,
possibly changed by small displacements $\{\epsilon_n\}$ lies on that surface:
\begin{equation}\label{eq:manifold} 
   F_q(\xx_n+\Eepsilon_n)=0, \qquad \forall q,n
\,.\end{equation}
The quantity $\sqrt{\av{\Eepsilon^2}}$ denotes the (root mean squared) average
error we make by approximating the points $\{\xx_n\}$ by the manifold ${\cal
  F}$. For a useful approximation we require the functions $F_q$ to be smooth
and the sequence $\{\epsilon_n\}$ to be small in the rms sense.

In a measurement, we can only obtain noisy data $\yy_n=\xx_n+\Eeta_n$, where
$\{\Eeta_n\}$ is some random contamination. The manifold ${\cal F}$ is not
known {\sl a priori} and has to be estimated from the data. By projecting the
points $\{\yy_n\}$ onto the estimated manifold ${\cal F}$ we can aim to recover
$\xx'_n=\xx_n+\Eepsilon_n$. If we can find a suitable manifold --- and carry
out the projections --- such that $\av{\Eepsilon^2}<\av{\Eeta^2}$, then we have
in fact reduced the observational error: The projected sequence is closer to the
true states $\{x_n\}$ than the noisy observations. 

In all the cases discussed in this paper, only a scalar measurement of the
system states is available: $s_n=s(\xx_n), \; n=1,\ldots,N$. A
multi-dimensional vector representation can be obtained by considering also
time delayed copies of the scalar sequence: $\Ss_n=(s_{n-(m-1)\tau},
s_{n-(m-2)\tau}, \ldots, s_n)$.  For deterministic dynamical systems, theorems
are available~\cite{Takens,embed} on the equivalence of the sequences
$\{\Ss_n\}$ and $\{\xx_n\}$.  Suppose a system is governed by deterministic
equations of motion in $m-1$--dimensional delay coordinate space. Then
Eq.(\ref{eq:determ}) becomes 
\begin{equation}\label{eq:delay}
   s_n - F(s_{n-(d-1)\tau},  \ldots, s_{(n-\tau)}) = 0, \qquad \forall n
\,.\end{equation}
This means that in $m$-dimensional embedding space, there exists an
$m-1$--dimensional manifold containing the signal. Noisy measurements of such
processes can be cleaned by imposing the relation~(\ref{eq:delay}) on the data.
In higher dimensional embeddings, more than one independent equations of the
form (\ref{eq:delay}) exist.  Still, these functions $F_q$ are unknown and have
to be approximated by a fit.

\begin{figure}
\centerline{\input{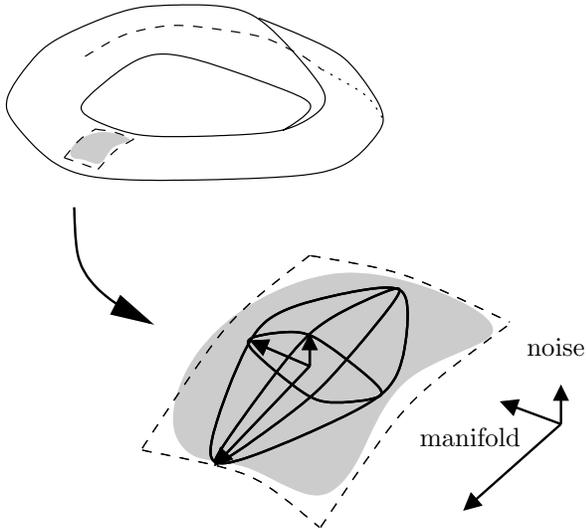}}
\caption[]{Illustration of the local projection scheme. For each point to be
  corrected, a neighbourhood is formed (grey shaded area), the point cloud in 
  which is then approximated by an ellipsoid. An approximately two-dimensional
  manifold embedded in a three-dimensional space could for example be cleaned
  by projecting onto the first two principal directions.\label{fig:noise}}
\end{figure}

In time series work, the most practical way to approximate data by a manifold
is by a locally linear representation. It should in principle be possible to
fit global nonlinear constraints $\hat{F}_q$ from data, but the problem is
complicated by the necessity to have $Q$ locally independent equations. In the
locally linear case this is achieved by establishing local principal
components. The derivation will not be repeated here, it is carried out for
example in Ref.~\cite{KantzSchreiber}. The resulting algorithm proceeds as
follows. In an embedding space of dimension $m$ we form delay vectors
$\Ss_n$. For each of these we construct small neighborhoods ${\cal U}_n$, so
that the neighbouring points are $\Ss_k, k\in{\cal U}_n$.  Within each
neighbourhood, we compute the local mean
\begin{equation}
   \overline{\Ss}^{(n)}=\frac{1}{|{\cal U}_n|}\sum_{k\in {\cal U}_n} \Ss_k
\end{equation}
and the $(m\times m)$ covariance matrix
\begin{equation}
   C_{ij}=\frac{1}{|{\cal U}_n|}\sum_{k\in {\cal U}_n}
   (\Ss_k)_i(\Ss_k)_j-\overline{\Ss}^{(n)}_i\overline{\Ss}^{(n)}_j
\,.\end{equation}
It has been found advantageous~\cite{on} to introduce a diagonal weight matrix
$R$ and define a transformed version of the covariance matrix
$\Gamma_{ij}=R_{ii} C_{ij}R_{jj}$ for the calculation of the principal
directions.  In order to penalise corrections based on the first and last
coordinates in the delay window one puts $R_{00}=R_{mm}=r$ where $r$ is large.
The other values on the diagonal of $R$ are~1.

The eigenvectors of the matrix $\Gamma_{ij}$ are the semi-axes of the best
approximating ellipsoid of the cloud of points. These are local versions of the
well known principal components, or singular vectors, see for example
Refs.~\cite{PC,BroomSVD}. If the clean data lives near a smooth manifold with
dimension $m_0<m$, and if the variance of the noise is sufficiently small for
the linearisation to be valid, then for the noisy data the covariance matrix
will have large eigenvalues spanning the smooth manifold and small eigenvalues
in all other directions. Of course, this is strictly true only if the
neighbourhoods are larger than the noise level. In practice, a tradeoff between
the clear definition of the noise directions and a good linear approximation
has to be balanced.

\begin{figure}[b]
\centerline{\input{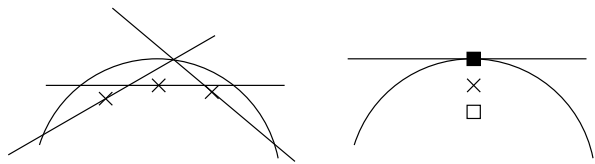}}
\caption[]{Local linear approximation to a one-dimensional
   curve. Left: approximations are not tangents but secants and all the centres
   of mass (crosses) of different neighbourhoods are shifted inward with
   respect to the curvature. Right: a tangent approximation is obtained by
   shifting the centre of mass outward with respect to the curvature. The open
   square denotes the average of the centres of mass of adjacent
   neighbourhoods, the filled square is the corrected centre of
   mass.\label{fig:tangent}}
\end{figure}

By projecting onto the subspace of large eigenvectors, we move the vector 
under consideration towards the manifold. The procedure is graphically
illustrated in Fig.~\ref{fig:noise}. The correction is done for each embedding
vector, resulting in a set of corrected vectors in embedding space.  Since each
element of the scalar time series occurs in $m$ different embedding vectors, we
finally have as many different suggested corrections, of which we simply take
the average. Therefore in embedding space the corrected vectors do not
precisely lie on the local subspaces but are only moved towards them.

If the local linear subspaces are determined in the way outlined above, they 
are not really tangent to  the curved manifold but
rather intersect with it, as illustrated in Fig.~\ref{fig:tangent}.
Therefore it is preferable to use a corrected centre of mass $\overline{\bf
s}^{(n)}$ given by 
\begin{equation}
   \overline{\overline{\bf s}}^{(n)} = 2\,\overline{\bf s}^{(n)}
      - {1 \over |{\cal U}_n|} \sum_{k\in {\cal U}_n} 
      \overline{\bf s}^{(k)} 
\end{equation} 
This correction prevents a bias towards corrections in the main direction of
curvature. If it is omitted, and if rather large neighbourhoods are used, 
the set of embedded data points in phase space may be subject to an overall
contraction on multiple iterations of the procedure.

A computer program that implements the scheme described in this paper is
included in the TISEAN software project and is publicly
available~\cite{tisean}.

\section{NUMERICAL EXAMPLE}
\begin{figure}
\centerline{\hspace{0.95cm}
\setlength{\unitlength}{0.1bp}
\begin{picture}(1800,1209)(0,0)
\special{psfile=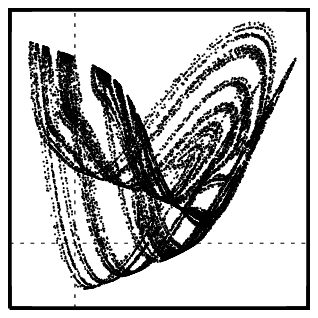 llx=0 lly=0 urx=360 ury=282 rwi=3600}
\put(1259,150){\makebox(0,0){1.8}}
\put(400,150){\makebox(0,0){-0.5}}
\put(350,1109){\makebox(0,0)[r]{1.8}}
\put(350,250){\makebox(0,0)[r]{-0.5}}
\end{picture}\hspace{-2.5cm}
\setlength{\unitlength}{0.1bp}
\begin{picture}(1800,1209)(0,0)
\special{psfile=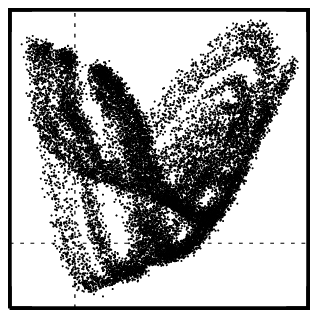 llx=0 lly=0 urx=360 ury=282 rwi=3600}
\put(1259,150){\makebox(0,0){1.8}}
\put(400,150){\makebox(0,0){-0.5}}
\put(350,1109){\makebox(0,0)[r]{1.8}}
\put(350,250){\makebox(0,0)[r]{-0.5}}
\end{picture}}
\centerline{\hspace{0.95cm}
\setlength{\unitlength}{0.1bp}
\begin{picture}(1800,1209)(0,0)
\special{psfile=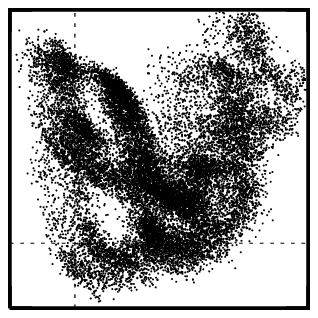 llx=0 lly=0 urx=360 ury=282 rwi=3600}
\put(1259,150){\makebox(0,0){1.8}}
\put(400,150){\makebox(0,0){-0.5}}
\put(350,1109){\makebox(0,0)[r]{1.8}}
\put(350,250){\makebox(0,0)[r]{-0.5}}
\end{picture}\hspace{-2.5cm}
\setlength{\unitlength}{0.1bp}
\begin{picture}(1800,1209)(0,0)
\special{psfile=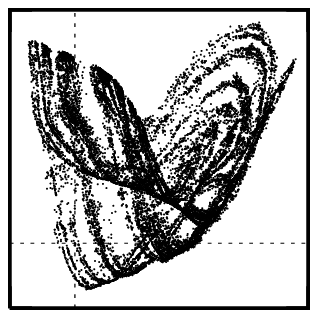 llx=0 lly=0 urx=360 ury=282 rwi=3600}
\put(1259,150){\makebox(0,0){1.8}}
\put(400,150){\makebox(0,0){-0.5}}
\put(350,1109){\makebox(0,0)[r]{1.8}}
\put(350,250){\makebox(0,0)[r]{-0.5}}
\end{picture}}
\caption[]{\label{fig:ikeda} Ikeda map example. Upper left: noise free data;
   right: data contaminated with 5\% additive Gaussian white noise. Lower left:
   data after low-pass filtering; right: data after nonlinear noise reduction.}
\end{figure}

\begin{figure}[t]
\centerline{
\setlength{\unitlength}{0.1bp}
\begin{picture}(1980,1080)(0,0)
\special{psfile=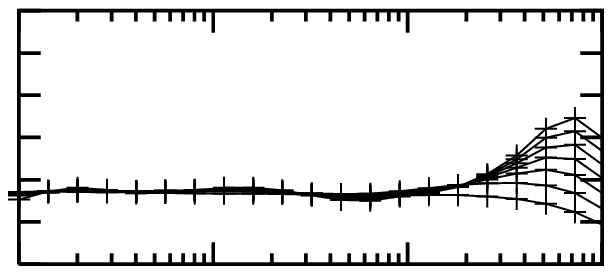 llx=0 lly=0 urx=396 ury=252 rwi=3960}
\put(1930,150){\makebox(0,0){1}}
\put(1370,150){\makebox(0,0){0.1}}
\put(810,150){\makebox(0,0){0.01}}
\put(250,150){\makebox(0,0){0.001}}
\put(200,980){\makebox(0,0)[r]{6}}
\put(200,858){\makebox(0,0)[r]{5}}
\put(200,737){\makebox(0,0)[r]{4}}
\put(200,615){\makebox(0,0)[r]{3}}
\put(200,493){\makebox(0,0)[r]{2}}
\put(200,372){\makebox(0,0)[r]{1}}
\put(200,250){\makebox(0,0)[r]{0}}
\end{picture}}
~\\[-25pt]
\centerline{
\setlength{\unitlength}{0.1bp}
\begin{picture}(1980,1080)(0,0)
\special{psfile=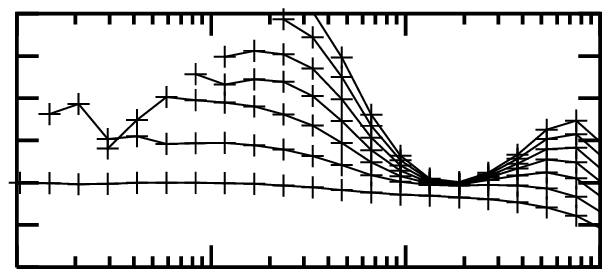 llx=0 lly=0 urx=396 ury=252 rwi=3960}
\put(1930,150){\makebox(0,0){1}}
\put(1370,150){\makebox(0,0){0.1}}
\put(810,150){\makebox(0,0){0.01}}
\put(250,150){\makebox(0,0){0.001}}
\put(200,980){\makebox(0,0)[r]{6}}
\put(200,858){\makebox(0,0)[r]{5}}
\put(200,737){\makebox(0,0)[r]{4}}
\put(200,615){\makebox(0,0)[r]{3}}
\put(200,493){\makebox(0,0)[r]{2}}
\put(200,372){\makebox(0,0)[r]{1}}
\put(200,250){\makebox(0,0)[r]{0}}
\end{picture}}
~\\[-25pt]
\centerline{
\setlength{\unitlength}{0.1bp}
\begin{picture}(1980,1080)(0,0)
\special{psfile=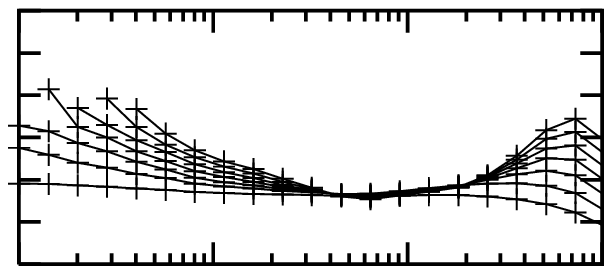 llx=0 lly=0 urx=396 ury=252 rwi=3960}
\put(1930,150){\makebox(0,0){1}}
\put(1370,150){\makebox(0,0){0.1}}
\put(810,150){\makebox(0,0){0.01}}
\put(250,150){\makebox(0,0){0.001}}
\put(200,980){\makebox(0,0)[r]{6}}
\put(200,858){\makebox(0,0)[r]{5}}
\put(200,737){\makebox(0,0)[r]{4}}
\put(200,615){\makebox(0,0)[r]{3}}
\put(200,493){\makebox(0,0)[r]{2}}
\put(200,372){\makebox(0,0)[r]{1}}
\put(200,250){\makebox(0,0)[r]{0}}
\end{picture}}
~\\[-25pt]
\caption[]{\label{fig:ikeda_dim} Local slopes of the correlation sum indicating
   fractal structure (Ikeda map example). Upper: noise free data. Middle: 
   data contaminated with 5\% additive Gaussian white noise. Lower: after
   nonlinear noise reduction. See text.}
\end{figure}

Let us show an example for the noise reduction capability of the algorithm for
deterministically chaotic time series. The Ikeda map is given by the formula
\begin{equation}
      z_{n+1} = 1 + 0.9 z_n \exp\left(0.4i- \frac{6i}{1+|z_n|^2}\right) 
\end{equation}
where $\{z_n\}$ is a sequence of complex numbers. Consider as a model time
series the sequence $x_n=\Re(z_n)$ given by the real parts of $z_n$.  In
Fig.~\ref{fig:ikeda}, the upper left panel shows a delay representation with
delay $\tau=1$ of the noise free sequence $\{x_n\}$. In order to mimic a
simplified experimental situation, the sequence $s_n=x_n+\eta_n$ has been
contaminated by adding Gaussian independent random numbers of a rms amplitude
of 5\% of that of the clean data, upper right panel.  The lower row shows two
different attempts on noise reduction. Left, a low pass filter has been applied
to suppress the highest 4\% of frequencies in the Nyquist interval. Since the
signal has still significant power at these frequencies, the filter is
inadequate and, in fact, severely distorts the phase space structure. In the
right panel, nonlinear noise reduction has been applied using embeddings in
$m=7$ dimensions and local projections onto $m-Q=3$ dimensions. The
neighbourhood size was taken to be 0.02 units, which is about the absolute noise
level added. The figure is the result of three iterations. The error was
reduced by a factor of 1.7 in terms of rms amplitude. Note however that the
data is probably much closer than that to a true trajectory of the Ikeda
system.

In Fig.~\ref{fig:ikeda_dim}, the effect of noise reduction on this fractal
attractor is demonstrated with the help of the Grassberger-Procaccia
correlation sum, $C(\epsilon)$, the fraction of pairs of points closer than
$\epsilon$ in delay coordinate space. As explained for example in
Ref.~\cite{KantzSchreiber}, we take the local slope in a double logarithmic
plot of $C(\epsilon)$ versus $\epsilon$ as an effective scale dependent scaling
exponent $d(\epsilon)=d \log C(\epsilon) / d \log \epsilon$. If a significant
plateau occurs and certain precautions have been taken, the plateau value of
$d(\epsilon)$ would estimate the correlation dimension of the attractor
underlying the data. Indeed, such a plateau can be seen for the noise free data
(upper panel) while it is rather small for the noisy data. After nonlinear
noise reduction (lower panel), the scaling is recovered down to much smaller
length scales.

\section{DISCUSSION}
With the emergence of experiments exhibiting deterministic chaos, nonlinear
filtering techniques became a necessity. The earliest approaches required
detailed understanding of the dynamics and the stability structure in phase
space. The phase space projection technique described in this paper goes back
to Ref.~\cite{on} but some earlier techniques are quite similarly
set-up~\cite{noiserev}. These techniques do not require a detailed model, or a
global fit of such a model, for the dynamics. Rather, the inhomogeneous
distribution in reconstructed phase space is enhanced by projecting onto local
linear subspaces. We have demonstrated in this paper that the method is
effective in reducing noise in an artificial, chaotic example. Applications to
real time series data will be discussed in Ref.~\cite{II} in this volume.


\begin{thebibliography}{10}
\bibitem{KantzSchreiber}
   H. Kantz and T. Schreiber, 
   {\em Nonlinear Time Series Analysis}.
   Cambridge Univ. Press, Cambridge (1997).

\bibitem{noiserev}  
E.~J.\ Kostelich and T.\ Schreiber,
   ``Noise reduction in chaotic time series data: A survey of common methods'',
   Phys.\ Rev.\ E 48 (1993) 1752.

\bibitem{on} P.\ Grassberger, R.\ Hegger, H.\ Kantz, C.\ Schaffrath, and
   T.\ Schreiber, 
   ``On noise reduction methods for chaotic data'',
   Chaos 3 (1993) 127.

\bibitem{II}
T.\ Schreiber and H.\ Kantz,
   ``Nonlinear projective filtering II: Application to real time series'',
   this volume (1998).

\bibitem{Takens}
F. Takens, 
   ``Detecting Strange Attractors in Turbulence'',
   Lecture Notes in Math.\ Vol.~898, Springer, New York (1981). 

\bibitem{embed} 
T.\ Sauer, J.\ Yorke, and M.\ Casdagli, 
   ``Embedology'',
   J.\ Stat.\ Phys.\ 65 (1991) 579.

\bibitem{PC}
I.~T. Jolliffe,
   ``Principal component analysis'',
   Springer, New York (1986).

\bibitem{BroomSVD}
D. Broomhead and G.~P. King,
   ``Extracting qualitative dynamics from experimental data'',
   Physica D 20 (1986) 217.

\bibitem{tisean}
The TISEAN software package is publicly available for download from 
   either
      {\tt http://\linebreak[1]%
      www.\linebreak[1]%
      mpipks-dresden.\linebreak[1]%
      mpg.\linebreak[1]%
      de/\linebreak[1]%
      $\tilde{~}$tsa/\linebreak[1]%
      TISEAN/\linebreak[1]%
      docs/\linebreak[1]%
      welcome.html}
      or
      {\tt http://\linebreak[1]%
      wptu38.\linebreak[1]%
      physik.\linebreak[1]%
      uni-\linebreak[1]%
      wuppertal.\linebreak[1]%
      de/\linebreak[1]%
      Chaos/\linebreak[1]%
      DOCS/\linebreak[1]%
      welcome.html}. 
\end{thebibliography}
\end{document}